\documentclass[conference]{IEEEtran}

\usepackage[dvips]{graphicx}
\usepackage[cmex10]{amsmath}
\usepackage{array}
\usepackage{mdwmath}
\usepackage{mdwtab}
\usepackage{booktabs}
\usepackage{multirow}
\usepackage{CJK}
\usepackage{subfig}
\usepackage{multirow}
\usepackage{float}
\usepackage{enumerate}
\usepackage{algorithm}
\usepackage{algpseudocode}

\makeatletter

\newcommand{\Rmnum}[1]{\expandafter\@slowromancap\romannumeral #1@}
\makeatother
\ifCLASSINFOpdf
  \else
\fi

\begin{document}

\title{Design of Link-Quality-prediction-based Software Defined Wireless Sensor Networks}

\author{
\IEEEauthorblockN{Liao Wenxing }
\IEEEauthorblockA{Information Engineering College\\
Shaoguan University \\
Shaoguan 512000, P.R.China\\
Email: liaowenxing@sgu.edu.cn}

\and
\IEEEauthorblockN{Shi Xiaofei }
\IEEEauthorblockA{Macau University of Science and Technology\\
Macau, P.R.China\\
Email: 604767519@qq.com}

}

\maketitle

\begin{abstract}
In wireless multi-hop networks, the instability of the wireless links leads to unstable networking. Even in the new designed Software Defined Wireless Sensor Networks (SDWSN), the similar problems exist. To further improve the stability of SDWSN, we introduce a Link Quality (LQ) prediction model into the SDWSN architecture. The prediction model is used to improve the stability between neighbor nodes, and thus the stability of wireless multi-hop routes. Simulation results show that the LQ prediction model can make reasonable corrections to the reality wireless link model, which can well address the restrictive nature of unstable links. As a result, by reducing the use of unstable wireless links, the stability of the SDWSN network improved.

\end{abstract}

\section{Introduction}
With the development of the Internet of Things (IoT), wireless sensor networks, as an important part of the IoT, also play an important role in an increasing number of fields. An important application scenario for wireless sensor networks is environmental monitoring. In such applications, as the scale of the network increases, the use of a single sink node to cover the monitoring area is increasingly unable to meet the application requirements.Therefore, most wireless sensor networks use a wireless multi-hop networking method. Although this networking method can greatly increase the scale of the network, the accumulation of errors in wireless multi-hop transmission makes networking and multi-hop data transmission extremely unstable.

Most wireless multi-hop networks, including wireless sensor networks, use a distributed routing strategy at the network layer. Our previous research [1-2] has demonstrated that, due to the instability of wireless transmission, regardless of whether active or passive routing strategies are used, the probability of routing success and network stability are greatly limited as network scale increases.

In view of this, researchers have tried to introduce the software-defined ideas in wired networks into wireless sensor networks to reduce or weaken the restrictive nature of unstable links in distributed networking. Some typical software-defined wireless sensor network architectures have been proposed [3-9]. In [3],the challenges and design requirements are discussed. There are two major architectures, two-layer architecture[4-5] and three-layer architecture[6]. In our previous study [7], the feasibility as well as the advantages and challenges of software-defined wireless sensor networks (SDWSN) has been analyzed in detail. Beside the design of SDWSN architecture, some researchers focus on the QoS of SDWSN [8] and new technologies(Edge computation [9] for example) in SDWSN.

Nevertheless, there are still limitations in the SDWSN architecture that cannot be ignored, mainly in the following two aspects. Firstly, in the centralized controlled networking approach, the distributed neighbor discovery process is inevitably necessary in order to obtain node connection relationships. The existence of weak links still leads to variable neighbor relationships, which leads to frequent topology repair[2]. Secondly, when the controller collects the neighbor information, it still adopts a broadcast method similar to passive routing protocols. Therefore,the weak links may be used, which leads to the failure of topology collection process.

The above problems are mainly caused by the uncertainty of wireless transmission. When two nodes are on their respective coverage edges, the link becomes unstable due to the temporal and spatial contingency of the wireless signals, which affects the stability of the neighbor relationship. In addition, the error accumulation characteristic of wireless multi-hop transmission further worsens this instability, which brings challenges in stable networking in SDWSN.

To address these issues, we propose an LQ prediction model-based SDWSN architecture, and implement the wireless multi-hop networking mechanism under this architecture. The LQ prediction model can predict the link conditions at the next broadcast period from the historical packet recption of wireless links, so that the existence of unstable links can be predicted in advance and the use of unstable links can be avoided during the networking strategy. Therefore, the application of LQ prediction models can reduce the possibility of using unstable links, which helps to improve the stability of wireless connections between neighbors and the reliability of multi-hop transmission.

At present, the researchers have proposed a number of link quality prediction methods, including traditional optimization algorithms, such as (EWMA)[11,24] and Machine learning-based algorithms, such as (fuzzy and svr)[12,13]. The historical values of link quality metrics such as Received Signal Strength Indicator(RSSI), Packet Receive Rate(PRR), Signal-Noise Ratio(SNR), and Link Quality Indicator(LQI) are used as the input parameters of prediction models. The output parameters mainly include the link quality over a period of time, the package reception for the next period and the received signal strength for the next period. The mainstream link quality prediction models are generally machine learning-based, and can be divided into two types, the off-line [15,18,21] and on-line [14,17,20] learning models. In [14] and [15], the authors propose an on-line learning model (TALENT) and an off-line learning model (4C) respectively. Both link quality prediction models focus on the improvement of packet delivery rate. The authors use the PRR, RSSI, SNR and LQI as the input parameters to train the models. In [16], the authors propose a link quality prediction protocol to deal with the Multi-robot Coordination Tasks. The RSSI, SNR and the neighbor information of nodes are applied in the proposed protocol. Authors in [10] summarize the link quality prediction applications related to power control. In [26], the authors focus on the measurements and characterization of link quality in energy-constrained 802.15.4 wireless sensor networks.

Consider the network model, most of the link prediction models are proposed for the static or quasi-static networks, such as MESH and WSN [14,15,17,20,22,23,24,25]. While for the networks with mobility, (MANET for example) [16,18,21], the additional network dynamic information is necessary for link quality prediction models. Otherwise, authors in [22,25] propose link quality prediction models for indoor and outdoor wireless sensor networks respectively.

A large number of prediction models achieve good prediction results in different application scenarios. However, due to the random property of wireless links, the models are not very accurate for predicting unstable links. Nevertheless, the use of the LQ prediction model is still able to improve the stability of the link well and reduce the instability region between nodes. This will be explained in detail in the following sections.

Due to the stabilizing effect of the LQ prediction model on wireless links, we introduced an LQ prediction mechanism into the original SDWSN architecture [7] and proposed an LQ prediction-based SDWSN  architecture. In this network architecture, we use the output results of  prediction model to restrict the usage of weak links and improve the stability and reliability of the network.

The main contributions of this paper are as follows:

\begin{enumerate}
\item  A wireless LQ prediction model is proposed and it is demonstrated that the LQ prediction model can effectively reduce the usage of unstable links during networking process.

\item  The design of the network architecture of SDWSN is improved, and the processes of neighbor discovery and topology discovery under the SDWSN architecture are optimized.

\item  The LQ prediction model is applied in SDWSN to effectively avoid the dynamic change of neighbors and improve the stability of the network.
\end{enumerate}

The rest of this paper is organized as follow. In section \Rmnum{2}, we design a machine learning-based wireless LQ prediction model and analyze the usefulness of the model in stabilizing links in detail. In section \Rmnum{3}, we design an SDWSN architecture based on the link prediction model and refine the design of the protocol flow. In Section \Rmnum{4}, we build a simulation platform to verify the improvement of network performance by engaging the LQ prediction model. Finally, we conclude this paper in section \Rmnum{5}.


\section{Design and analysis of wireless link prediction model }

\subsection{The structure of LQ prediction model}
In this paper, we propose a machine learning-based link prediction model. The structure of the proposed model is as shown in Fig 1.

\begin{figure}[h]
\centering
\vspace{0.2cm}
\includegraphics[width=3.4 in]{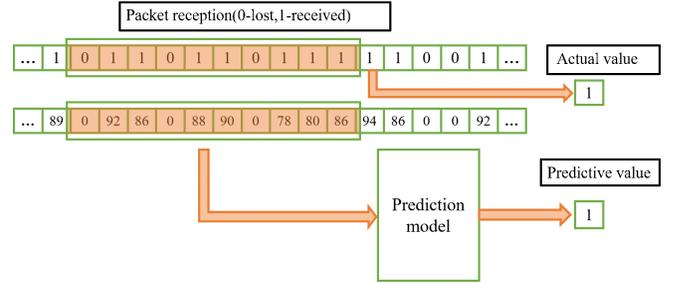}
\centering{\caption{The Structure of LQ prediction model}\label{}}
\end{figure}

The inputs to the model combine physical parameters with logical parameters. Suppose a HELLO packet is sent periodically between two nodes, by using the physical parameters and the logical parameters from the $1_{st}$ to the $k_{th}$ broadcast period, we can predict the reception of the HELLO packets in the $(k+1)_{th}$ period. In this paper, the physical parameters use the RSSI value of the HELLO packets and the logical parameters are the reception of the historical HELLO packets.

First, $i = \{RSSI,RECV \}$ is defined as the input parameter of the model and the reception in the $(k+1)_{th}$ broadcast period is defined as the output of the model. We obtain these parameters based on the packet reception during the previous $k$ broadcast periods between two nodes in a certain link.

$RSSI_i$ indicates the signal strength of the $i_{th}$ HELLO packet and $RECV_i$ indicates the reception of the $i_{th}$ HELLO packet, where $RECV_i=1$ indicates that the $i_{th}$ HELLO packet was successfully received and $RECV_i=0$ indicates that the $i_{th}$ HELLO packet was lost. In the prediction model, we use the $RECV$ and $RSSI$ of the previous $k$ HELLO packets as input parameters and the reception of the $(k+1)_{th}$ HELLO packet as the predicted output.

Then we build a prediction model based on machine learning, mainly using XGboost, neural networks, etc., and train the model with a large number of input samples. The trained model is able to take in the input feature parameters to accurately predict the reception of packets in the next period. Comparing the predicted output with the actual reception in the next period, we verify the performance of the prediction model.

In practice, the trained model is deployed at the data link layer. When a node receives a packet from its neighbor node, it obtains the historical reception of that node and makes a prediction. If the predicted output is received, the packet is handed over to the network layer for processing. Otherwise, the packet is discarded, indicating that it is a packet that should not be received for processing.

\subsection{The impact of the wireless link model on link stability}

A simulation platform has been built to simulate the process of data transmission between wireless nodes. Through the simulation platform, we obtain simulation data for different environments and use them for training and testing. Afterwards, the trained model is deployed to the simulation platform to verify the stability of the wireless LQ prediction model for the link. The data transfer process on the simulation platform is shown in Fig 2.

\begin{figure}[h]
\centering
\vspace{0.2cm}
\includegraphics[width=3.4 in]{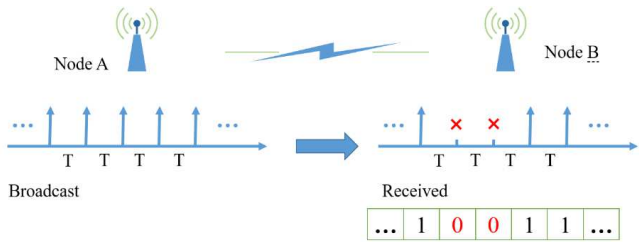}
\centering{\caption{HELLO packet broadcast process}\label{}}
\end{figure}

\subsubsection{Channel model}

Due to the fading nature of wireless transmission, packet loss may occur when data is transmitted between two nodes. Therefore, we use a log-normal shadow fading based wireless channel model on our simulation platform to simulate the wireless transmission environment between two nodes.

Let $x$ denotes the distance between nodes $i$ and $j$, and the packet delivery rate between these two nodes is given by $P(x)$. The effective communication radius is $r_0$. According to the log-normal shadow fading channel model[1], the packet delivery rate $P(x)$ is given by

\begin{equation}
p(x) = \frac{1}{2} - \frac{1}{2}erf\left( {\frac{{10\alpha }}{{\sqrt 2 \sigma }}{{\log }_{10}}\frac{x}{{{r_0}}}dB} \right)
\end{equation}
where $\alpha $ is the path loss exponent and $\sigma $ denotes the standard deviation, ${r_0}$ represents the effective communication radius, which is the distance when the point-to-point packet delivery rate drops to 50\%. Then we have

\begin{equation}
{r_0} = {10^{\frac{{{\beta _{th}}}}{{10\alpha }}}}
\end{equation}

\begin{equation}
{\beta _{th}} = 10{\log _{10}}\frac{{{p_t}}}{{{p_{r.th}}}}dB
\end{equation}
where ${\beta _{th}}$ is the signal attenuation threshold and ${p_{r,th}}$ is the threshold of received signal strength, ${p_t}$ is the sending signal strength. The curve of $P(x)$ is shown in fig 3. In this paper, we set $\alpha=3 $, $\sigma=4 $ and ${\beta _{th}}=66dB$.

\begin{figure}[h]
\centering
\vspace{0.2cm}
\includegraphics[width=3.4 in]{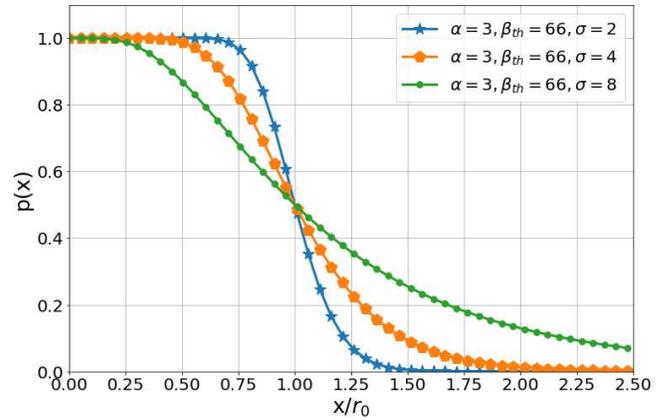}
\centering{\caption{Delivery rate of two nodes with distance x}\label{}}
\end{figure}

\subsubsection{Performance of the prediction model}

In order to understand the impact of the wireless LQ prediction model on unstable links, we have analyzed the performance of the prediction model in the following aspects.

\begin{itemize}
\item  \textbf{Accuracy}
\end{itemize}

We use the node's historic information of reception for the first K periods as parameters and use common machine learning models such as Regression model, SVM, Random Forest, XGBoost and Neural networks for link prediction. To better measure the performance of the LQ prediction models, we analyze the performance metrics obtained for each model, such as F1 values, ACC and recall. The results are as shown in Table \Rmnum{1}. Moreover, we performed a large number of simulations at different distances to obtain samples with distances between nodes uniformly distributed in the range of 0 to $2r_0$.

\begin{table}[!t]
\renewcommand{\arraystretch}{1.3}
\caption{Performance of LQ prediction models}
\label{table_example}
\centering
\begin{tabular}{|c||c||c||c||c|}
\hline
\textbf{Model}  & \textbf{ACC} & \textbf{F1} & \textbf{Precision} & \textbf{Recall} \\
\hline
Regression model & 0.8916 & 0.8627 & 0.8864 & 0.8402 \\
\hline
Decision Trees & 0.8411 & 0.8047 & 0.8015 & 0.8078 \\
\hline
Random Forest & 0.8890 & 0.8590 & 0.8851 & 0.8343 \\
\hline
CBDT & 0.8904 & 0.8606 & 0.8873 & 0.8355 \\
\hline
XGboost & 0.8904 & 0.8609 & 0.8871 & 0.8369 \\
\hline
SVM & 0.8897 & 0.8598 & 0.8863 & 0.8349 \\
\hline
Neural networks & 0.8915 & 0.8622 & 0.8885 & 0.8374\\
\hline
\end{tabular}
\end{table}

From Table \Rmnum{1}, we obtain that with suitable input and output parameters, the LQ prediction models based on different machine learning algorithms do not differ significantly and all achieve good prediction accuracy.

In fact, accuracy is of little importance when it comes to wireless link prediction. Chance is inevitable in actual transmission due to the randomness of the transmission process. Even if the distance between the two nodes is very close or far, there will be packet loss or packet reception. As a result, the output of the prediction model is also somewhat stochastic, which in turn leads to less accurate predictions. In practice, even though the prediction model is not very accurate, it has a great advantage in terms of link stability.

\begin{itemize}
\item  \textbf{Comparison of measured and predicted data}
\end{itemize}

To study the performance of the prediction model in unstable links, we compare the predicted and actual outputs for the three cases of inter-node distances of $0.8r_0$, $r_0$ and $1.2r_0$ respectively. Fig 4 to 6 show a comparison of the predicted and actual outputs of the nodes over a continuous period of time, with inter-node distances of $0.8r_0$, $r_0$ and $1.2r_0$ respectively.

\begin{figure}[h]
\centering
\vspace{0.2cm}
\includegraphics[width=3.4 in]{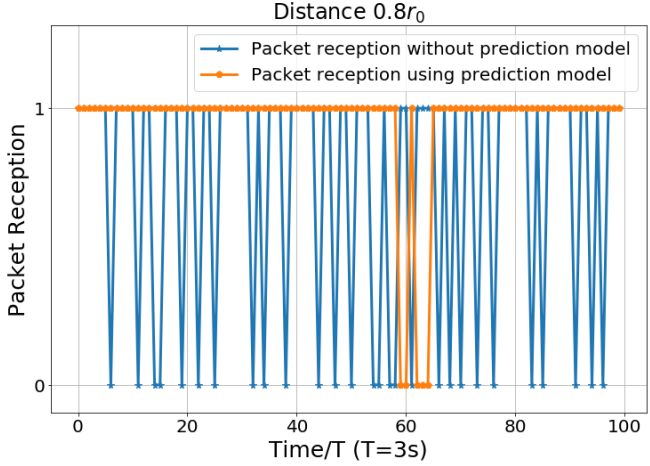}
\centering{\caption{The comparison of predicted and actual outputs ($0.8r_0$)}\label{}}
\end{figure}

\begin{figure}[h]
\centering
\vspace{0.2cm}
\includegraphics[width=3.4 in]{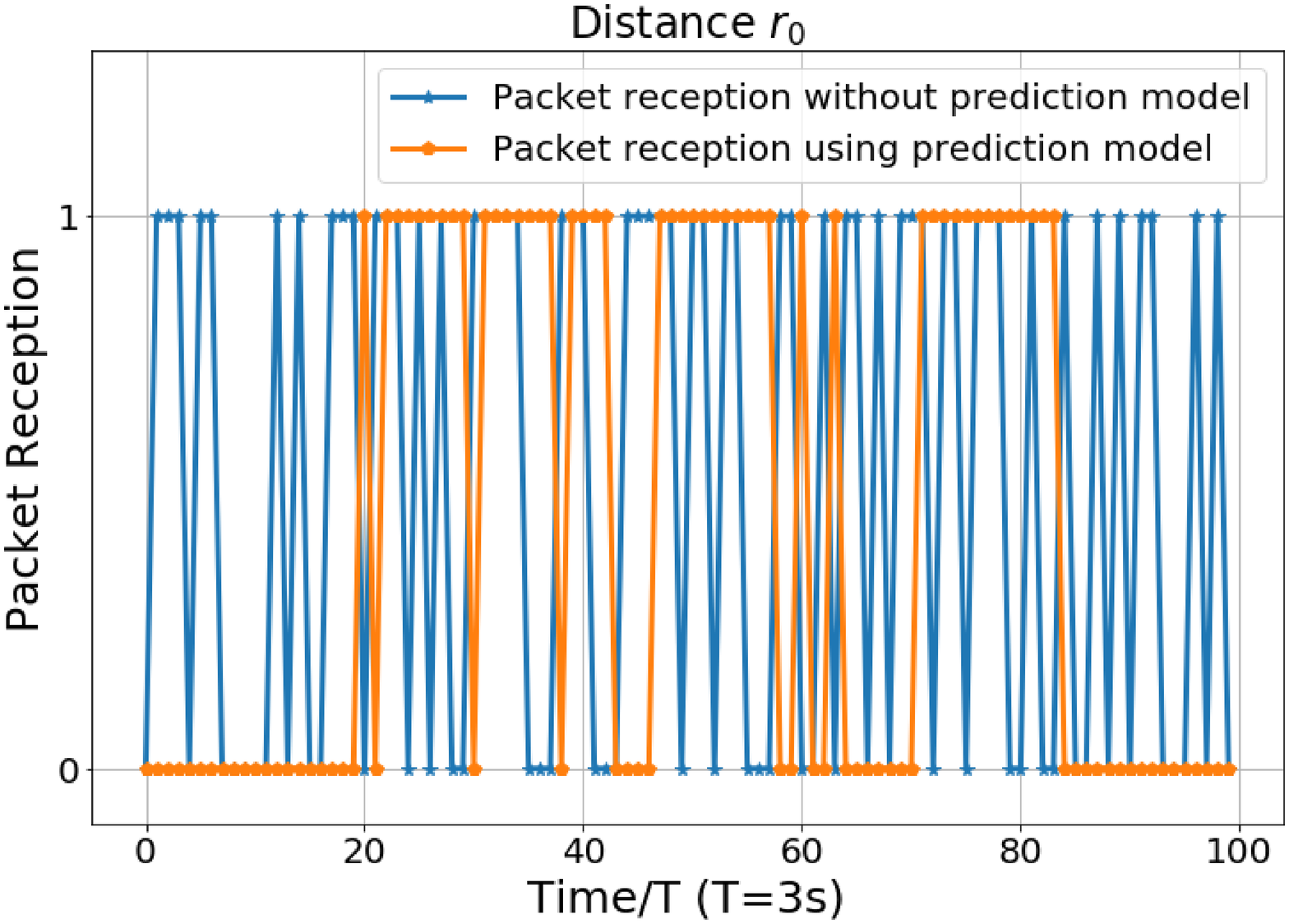}
\centering{\caption{The comparison of predicted and actual outputs ($r_0$)}\label{}}
\end{figure}

\begin{figure}[h]
\centering
\vspace{0.2cm}
\includegraphics[width=3.4 in]{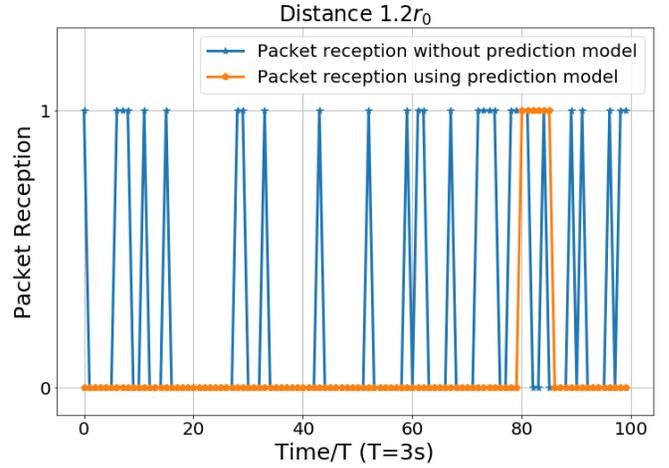}
\centering{\caption{The comparison of predicted and actual outputs ($1.2r_0$)}\label{}}
\end{figure}

As can be seen from the figures, if the link between two nodes is unstable, the packets are sent and received in a highly random situation without an LQ prediction model. After using the prediction model, the output of the predictor is relatively stable and does not show dramatic fluctuations.

\begin{itemize}
\item  \textbf{Modification of the channel by the prediction model}
\end{itemize}

We deployed the prediction model on two static nodes at distance x, ran it for a period of time and counted the accuracy of the prediction outputs. The prediction accuracy at different distances x is shown in Fig 7.

\begin{figure}[h]
\centering
\vspace{0.2cm}
\includegraphics[width=3.4 in]{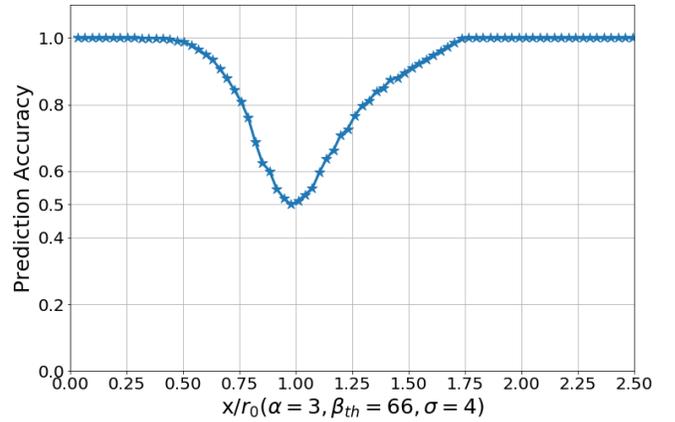}
\centering{\caption{The prediction accuracy at different distances x}\label{}}
\end{figure}

As can be seen from Fig 7, the accuracy of the predicted output varies with the distance. Prediction accuracy is worst when the distance of two nodes are near $r_0$, and highest when they are very close or far apart. This is because the link between two nodes is the most stable when they are very close or far apart.  However, when the distance is around $r_0$, the stochasticity is strong and the output of the predictor is almost random. In fact, machine learning algorithms cannot solve the problem of predicting purely random events. This also illustrates that there is little significance in pursuing accuracy in link prediction applications, so we need to approach link prediction models from a different perspective.

We incorporate the LQ prediction model into the actual link model, using the predicted output as the actual output, to investigate what happens to the link model after the prediction model is applied. A comparison of the variation of delivery rate before and after using the prediction model is shown in Fig 8.

\begin{figure}[h]
\centering
\vspace{0.2cm}
\includegraphics[width=3.4 in]{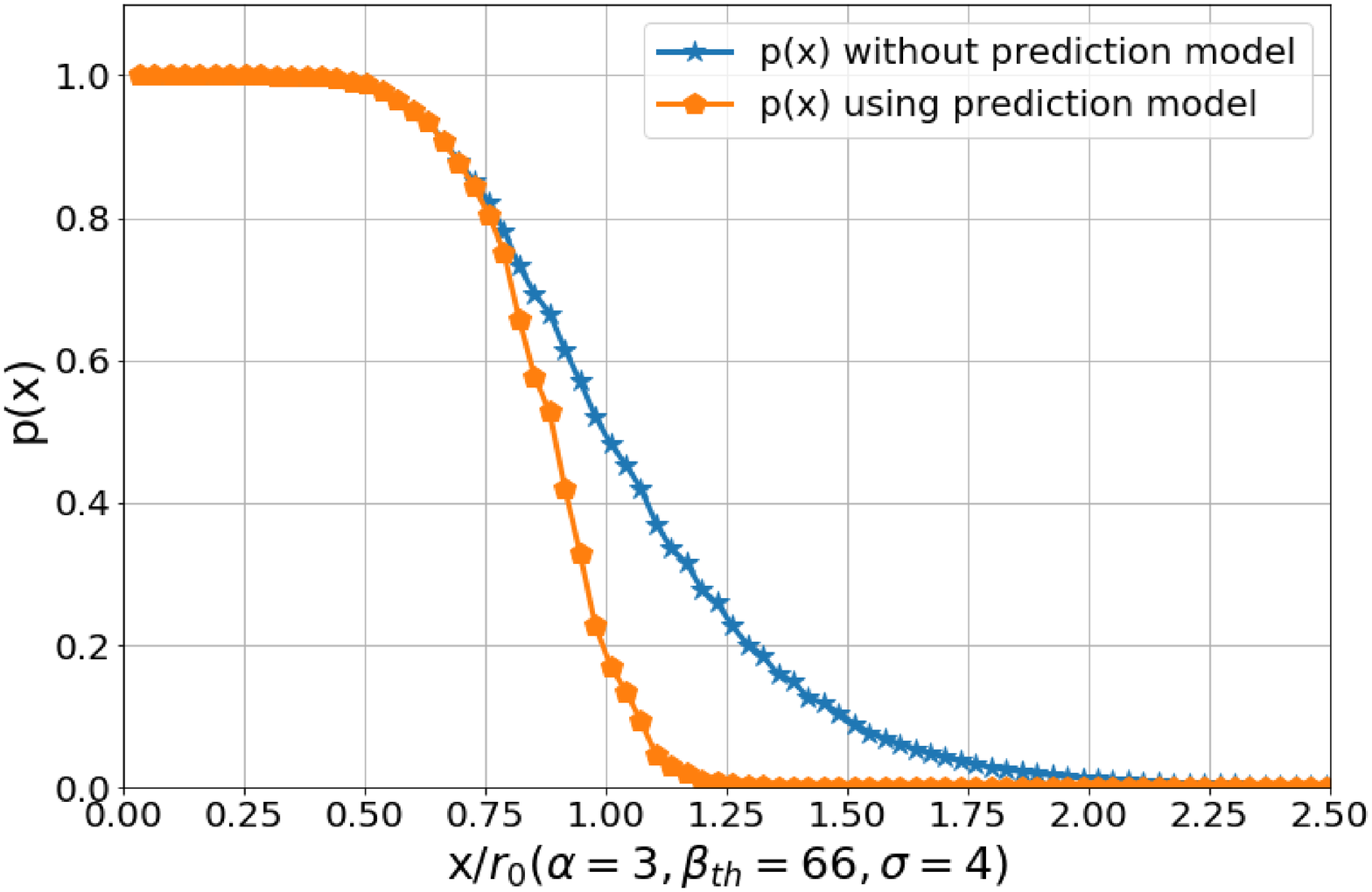}
\centering{\caption{The comparison of delivery rate of two nodes with distance x}\label{}}
\end{figure}

As can be seen from Fig 8, the delivery rate curve becomes steeper with the use of the prediction model. This corresponds to the fact that we have reduced the effective communication radius of the node, but the stability of the wireless link around the node's coverage radius has improved considerably. If we define a wireless link with a delivery rate between $30\%$ and $70\%$ as an unstable link, with the use of wireless LQ prediction model, it is clear that the range of unstable areas has been reduced. From previous theoretical studies, we know that the reason for the instability of wireless multi-hop networks is due to the presence of an instability region, the larger this region, the more unstable the network. The use of link prediction is a good way of reducing the presence of unstable links and avoiding the selection of unstable connections as data transmission channels in practical applications, thereby avoiding transmission failures.

In view of the above analysis, the use of a LQ prediction model in the SDWSN network architecture can be a good solution to the situation of unstable links and avoid the use of unstable links that cause network failure and massive packet loss in practical applications.

\section{Design of link prediction model based SDWSN architecture}
In this paper, we propose an SDWSN architecture based on a link prediction model. Improving on the existing SDN-based wireless sensor network architecture, a link prediction model is proposed to improve the stability of the point-to-point link and hence the overall network performance.

\subsection{Restrictive analysis of SDWSN}

A software-defined wireless sensor network architecture is proposed in the [7], in which the applicability of software-defined ideas in wireless sensor networks is analyzed in detail and a networking strategy based on the idea of centralized control is designed. In this strategy, the controller node is responsible for the collection of the whole network topology and the generation and distribution of the flow table of the whole network nodes. The ordinary sensor nodes are only responsible for maintaining their own neighbor tables, in order to avoid the limitations brought by the distributed networking algorithm and improve the stability of networking. Despite the use of the idea of centralized control in the SDWSN network architecture, however, as there is no absolutely reliable secure channel for the multi-hop transmission of control information, it is still inevitable that algorithms such as distributed topology discovery are used. A a result, the disadvantages of a distributed large-scale wireless multi-hop network are unavoidable. This is reflected in two main aspects.

Firstly, the distributed neighbor discovery algorithm results in a dynamically changing neighbor table. In wireless multi-hop networks, neighbor discovery is a prerequisite for topology discovery and an essential networking phase. Nodes usually perform neighbor discovery by sending HELLO packets periodically. However, due to the instability of wireless transmission, there is a certain probability of packet loss. A strong connection may experience occasional packet loss, while a weak connection may also experience occasional packet reception. Even if some strategies are used for optimization, such as receiving or losing multiple packets in a row before determining a neighbor change, the neighbor relationship between any two nodes is still in dynamic changes, as has been demonstrated theoretically in [2]. It is this dynamic neighbor change that makes ordinary nodes initiate topology repair frequently, which on one hand increases the networking overhead and on the other hand may bring about network-wide asynchrony.

Secondly, the distributed topology-related algorithms bring about the loss of control information. Since there is no absolutely reliable and secure channel, topology control information has to be transmitted in a wireless multi-hop approach. This process is similar to the routing discovery process in passive routing protocols. Due to the error accumulation nature of wireless multi-hop transmissions and the presence of weak links, the delivery rate of multi-hop transmissions decreases dramatically with increasing network scale. Although the topology collection process in SDWSN does not require specifying the destination node for routing control information in the same way as the route discovery process of passive routing protocols, (or even the return of topology information by all nodes), the process of returning topology information is similar to the process of returning routing responses. In the SDWSN architecture, the control information may also be relayed in the downlink direction using weak links, and therefore also faces a dramatic deterioration in uplink transmission performance. This situation is also present in the flow table rule distribution process, and although this process can be improved by controller-related algorithms, the loss of control information is inevitable due to the presence of weak links in the neighbor discovery process.

The above drawbacks are ultimately due to the inability to guarantee the reliability of each link segment in a wireless multi-hop network and the inability to avoid the existence of weak links. In view of this, we have designed an SDWSN architecture based on LQ prediction to guarantee the reliability of each wireless link segment by link quality prediction, to minimize the use of unstable links and to improve reliability at the control level. The prediction results are also used at the controller side for flow table generation to improve transmission reliability at the data level.

\subsection{Design of link prediction model based SDWSN}

In view of the above flawed analysis, the use of distributed algorithms and wireless multi-hop paths cannot be avoided in SDWSN networks, and this paper introduces link prediction mechanisms into the design of SDWSN architectures.

In the previous section, we designed a machine learning-based model for wireless LQ prediction model. The model uses a large number of historical transmissions between two nodes as input to train the model, and then uses the trained model to predict the next moment of transmissions, as shown in Fig 1. In the previous section, we have verified the superiority of the wireless LQ prediction model in stabilizing links. Therefore, we have introduced an LQ prediction mechanism in the SDWSN architecture to improve the stability of inter-neighbor connectivity and the reliability of multi-hop transmission of control information by reducing the use of weak links. In this way, we address or weaken the restrictive nature of unstable links in topology management and neighbor management.

\begin{figure}[h]
\centering
\vspace{0.2cm}
\includegraphics[width=3.4 in]{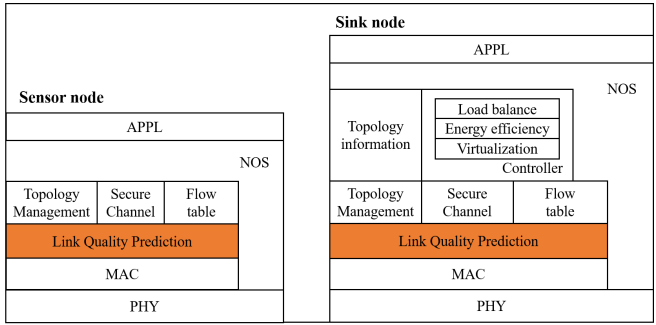}
\centering{\caption{Delivery rate of two nodes with distance x}\label{}}
\end{figure}

With the use of the LQ prediction mechanism, we propose a novel SDWSN architecture based on the link prediction model, as shown in Fig 9, which improves on the SDWSN architecture proposed in [7]. Similarly, there are two roles in this network architecture, the controller node and the sensor node, where the controller node is responsible for global network control, including global topology management and flow table rule management, and the sensor node is responsible for local topology management. The control level of SDWSN is implemented by two main processes, namely the neighbor management process and the topology management process.

\begin{itemize}
\item  Neighbor management process: the network-wide nodes (including the controller) perform local topology discovery, i.e. neighbor discovery, and collect neighbor information periodically.
\item  Topology management process: Controller-led, responsible for topology discovery, collection and maintenance, etc. There are three main operational phases as follows.
    \begin{itemize}
    \item Topology discovery phase. After the neighbor information of the sensing nodes has stabilized, the network enters the topology acquisition phase, when the controller node initiates the topology acquisition process to obtain the topology information of the whole network.
    \item Flow table deliver phase. The controller generates the optimal flow table for each node through the topology information collected from the whole network and delivers it to the sensor nodes in turn.
    \item Topology maintenance phase. All nodes operate stably, and when sensor nodes detect the changes in their neighbor tables, they started a topology repair process to report the changes to the controller.
    \end{itemize}
\end{itemize}

Inevitably, at the control level of the SDWSN, both neighbor management and topology management use a distributed approach, and the quality of the neighbor relationships obtained in neighbor discovery largely determines the performance of topology discovery. Therefore, we have designed the SDWSN architecture based on an LQ prediction model to maximize the reliability of each link, and thus the reliability of multi-hop transmission of topology control information.

\begin{itemize}
\item  \textbf{Neighbor management}
\end{itemize}

Before the controller performs topology collection, the sensor nodes need to perform neighbor discovery and refine their neighbor tables. All nodes, including the controller broadcast $HELLO_{RQ}$ packets periodically. After receiving the $HELLO_{RQ}$ packet, the nodes determine whether the $HELLO_{RQ}$ packet is processed by the LQ prediction model and modify the neighbor list of this node according to the processing result. In order to further ensure the stability of neighbor relationship, the link will be considered available after continuously receive M $HELLO_{RQ}$ packets. Additionally, the link will be considered unavailable after continuously last K $HELLO_{RQ}$ packets. The neighbor discovery process is shown in algorithm 1.

\begin{algorithm}
\floatname{algorithm}{Algorithm}
\renewcommand{\algorithmicrequire}{\textbf{Initialization:}}
\renewcommand{\algorithmicensure}{\textbf{Main\ loop:}}
        \caption{Process of Neighbor Discovery}
        \begin{algorithmic}[1]
            \Require Set M, K, T and N (N nodes in the network). Set recv=[0,0,...,0],lost=[0,0,...,0] and last\_recv\_time=[0,0,...,0] (N elements in recv, lost and last\_recv\_time list). Start a timer $t$ to periodically broadcast HELLO\_RQ.  Then start the neighbor discovery process.
            \Ensure
            \While {1}
            \State Wait until an interrupt occur
            \If {A HELLO\_RQ packet from node i is received}
                \State Get the historical reception of HELLO\_RQ packets of node i
                \State Get the prediction result of current HELLO\_RQ packet pre\_i
                \If {pre\_i == True}
                    \State recv[i]+=1
                    \State lost[i]=0
                    \State last\_recv\_time[i]=current time
                    \If {recv[i]$\>$=M and i not in neighbor list}
                        \State Add node i into neighbor list
                    \EndIf
                \Else
                    \State Drop this HELLO\_RQ packet
                \EndIf
            \ElsIf {Timer $t$ runs out}
                \State Broadcast a HELLO\_RQ packet
                \For {each $i \in N nodes$}
                    \If {current time $\>$ last\_recv\_time[i] + T}
                        \State lost[i]+=1
                        \State recv[i]=0
                        \If {lost[i]$\>$=K and i in neighbor list}
                            \State Move node i out of the neighbor list
                        \EndIf
                    \EndIf
                \EndFor
                \State Reset the timer $t$
            \EndIf
            \EndWhile
        \end{algorithmic}
\end{algorithm}

\begin{itemize}
\item  \textbf{Topology management}
\end{itemize}

When the network has been running for some time, the neighbor tables of all nodes are stable. At this time, the controller initiates the topology collection command and broadcasts the $Topology_{RQ}$ packet for network-wide topology collection. After completing the topology information collection, the controller enters the topology stable state, generates flow table rules for each node and sends them down to each node hop by hop, completing the networking function. During the stable state of the network, once a sensor node detects a neighbor change, the changed local topology is uploaded to the controller. Similarly, during all control commands issued, the sensor node uses a link prediction model to determine whether or not to process the control command. The specific topology management algorithms at the controller side and at the sensor side are as follows.

\begin{algorithm}
\floatname{algorithm}{Algorithm}
\renewcommand{\algorithmicrequire}{\textbf{Initialization:}}
\renewcommand{\algorithmicensure}{\textbf{Main\ loop:}}
        \caption{Process of Topology Management at Controller}
        \begin{algorithmic}[1]
            \Require Set timer $t1$. Then start the topology management process.
            \Ensure
            \While {1}
            \State Wait until an interrupt occur
            \If {Timer $t1$ runs out}
                \State Broadcast Topology\_RQ packet
                \State Set Timer $t2$
                \State Wait for Topology\_RP packets from sensors
                \While {1}
                    \If {Receive Topology\_RP from node i}
                        \State Get the neighbor list of node i
                        \State Add node i into global topology
                    \ElsIf {Timer $t2$ runs out}
                        \State Generate and delivery the flow table
                    \EndIf
                \EndWhile
            \EndIf
            \EndWhile
        \end{algorithmic}
\end{algorithm}

\begin{algorithm}
\floatname{algorithm}{Algorithm}
\renewcommand{\algorithmicrequire}{\textbf{Initialization:}}
\renewcommand{\algorithmicensure}{\textbf{Main\ loop:}}
        \caption{Process of Topology Management at sensors}
        \begin{algorithmic}[1]
            \Require Start the topology management process.
            \Ensure
            \While {1}
            \State Wait until an interrupt occur
            \If {Receive a new Topology\_RQ from node i}
                \If {The source of this Topology\_RQ in neighbor list}
                    \State Get the historical reception of HELLO\_RQ packets of node i
                    \State Get the prediction result of Topology\_RQ packet pre\_i
                    \If {pre\_i == True}
                        \State Generate a Topology\_RP packet
                        \State Send the Topology\_RP to the controller
                    \Else
                        \State Drop this Topology\_RQ packet
                    \EndIf
                \Else
                    \State Drop this Topology\_RQ packet
                \EndIf
            \ElsIf {Receive a duplicate Topology\_RQ packet}
                \State Drop this Topology\_RQ packet
            \EndIf
            \EndWhile
        \end{algorithmic}
\end{algorithm}

After the local topology discovery and global topology discovery, the network enters a stable state. At this point, the transmission path of data is completely determined by the flow table rules, and all nodes do not need to know the location of the destination node of the data, and complete the forwarding of data according to the flow table, realizing the separation of the control level and data level. Due to the use of the LQ prediction model, the unstable links can be well excluded by the prediction model in the reception of control information at each stage of the network, effectively improving the stability of the local topology and the global topology.

\section{Performance analysis }
To verify the performance of the LQ prediction model-based SDWSN architecture, we built a python-based multi-threaded distributed SDWSN simulation platform.

The simulation parameters are shown in Table \Rmnum{2}

\begin{table}[!t]
\renewcommand{\arraystretch}{1.3}
\caption{Introductions of some important symbols}
\label{table_example}
\centering
\begin{tabular}{|c||c|}
\hline
\textbf{Parameter}  & \textbf{Value} \\
\hline
Path loss exponent $\alpha $ & 3 \\
\hline
Standard deviation $\sigma $ & 4 \\
\hline
Signal attenuation threshold ${\beta _{th}}$ & 66dB\\
\hline
Effective communication radius $r_0$ & 150m\\
\hline
M & 1/2/3\\
\hline
K & 1/2/3\\
\hline
Node density $\rho$ & $(4-28)*10^{-5}nodes/m^2$\\
\hline
Network size & 500*500$m^2$\\
\hline
\end{tabular}
\end{table}

The nodes are evenly distributed in the simulation area and the simulation platform assigns each node an IP address, 127.0.0.ID, where ID is the number of the node. At the same time, the simulation platform assigns two threads to each node for neighbor management and topology management respectively. The controller node is located at the center of the simulation area and is responsible for network-wide control. The distribution of nodes is shown in Fig 10.

\begin{figure}[h]
\centering
\vspace{0.2cm}
\includegraphics[width=3.4 in]{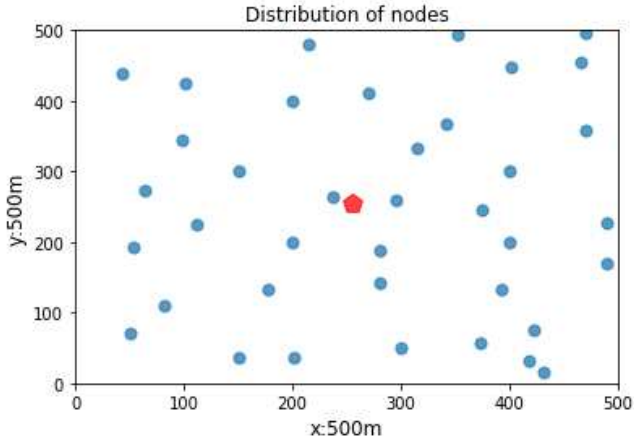}
\centering{\caption{Distribution of nodes}\label{}}
\end{figure}

To verify the change in network performance after the use of the LQ prediction model, we deploy a trained LQ prediction model designed in Section \Rmnum{2} for each node. When a node receives a packet from another node, the historical acceptance of that node is used as input to the model for link quality prediction. If the prediction result is 1, the packet is processed further, otherwise the packet is discarded.

\subsection{Analysis of local topology}

We analyze the local topology performance in two main ways. First, we analyze the link stability time between two given nodes A and B. Second, we analyze the neighbor stability time for a given node A at different node densities.

\begin{itemize}
\item  \textbf{Link stable duration}
\end{itemize}

Given a distance d between two nodes A and B, neighbor discovery is performed between the two nodes by broadcasting $HELLO_{RQ}$ packets. When node A receives M consecutive $HELLO_{RQ}$ packets from node B, a connection to node B is established, indicating that link $B->A$ exists. When node A does not receive $HELLO_{RQ}$ packets from node B for K consecutive broadcast periods, it disconnects from node B, indicating that link $B->A$ does not exist. The converse is also true.

\begin{figure}[h]
\centering
\vspace{0.2cm}
\includegraphics[width=3.4 in]{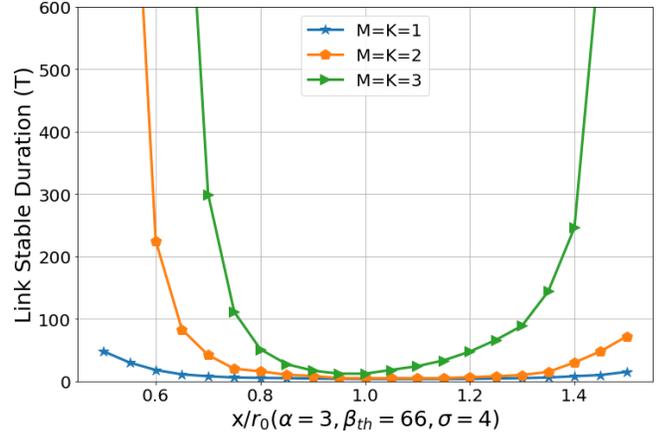}
\centering{\caption{The link stable durations between node A and B (without prediction)}\label{}}
\end{figure}

\begin{figure}[h]
\centering
\vspace{0.2cm}
\includegraphics[width=3.4 in]{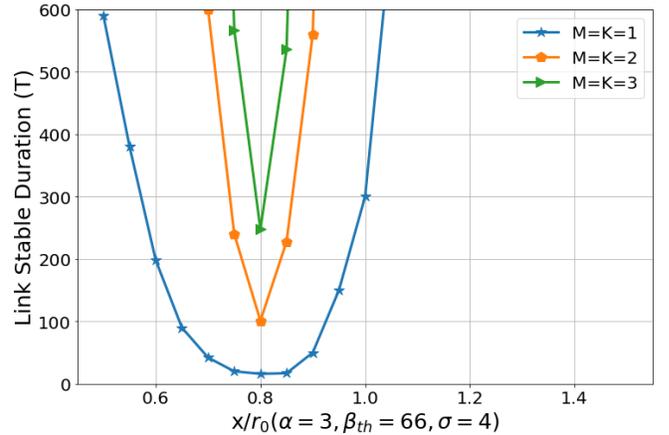}
\centering{\caption{The link stable durations between node A and B (prediction)}\label{}}
\end{figure}

Fig 11 and Fig 12 show the link stable durations without and with the prediction model respectively. It can be seen from Fig 11 that when the distance between two nodes is small or large, the link is able to remain stable for a long time. When the distance is around $r_0$, the link becomes unstable. As the values of M and K increase, the stable duration increases. Fig 12 shows the link stable durations after using the LQ prediction model. As can be seen from the figure, there is a significant improvement in link stability after using the LQ prediction model. The distance between two nodes is around $0.8r_0$, where the link stability is the worst. Combined with Fig 8, it can be seen that after the use of the LQ prediction model, the distance between two nodes with a delivery rate of around $50\%$ is around $0.8r_0$, where the link is the least stable. Nevertheless, the link stability is greatly improved relative to the case where no LQ prediction model is used. Again, the link stable duration increases as the values of M and K increase.

\begin{itemize}
\item  \textbf{Neighbor stability duration}
\end{itemize}

For a given node A, the number of potential neighbors of node A increases as the node density of the simulation platform increases. A link change between any segment of neighbors results in a local topology change of node A. We define the time for which the neighbor table of node A remains stable as the neighbor stable duration. Neighbor stability in SDWSN with and without the LQ prediction model is studied separately.

\begin{figure}[h]
\centering
\vspace{0.2cm}
\includegraphics[width=3.4 in]{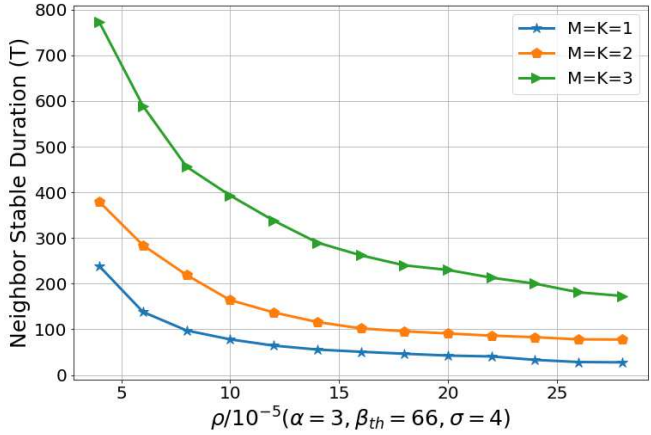}
\centering{\caption{The neighbor stable durations of node A (prediction)}\label{}}
\end{figure}

\begin{figure}[h]
\centering
\vspace{0.2cm}
\includegraphics[width=3.4 in]{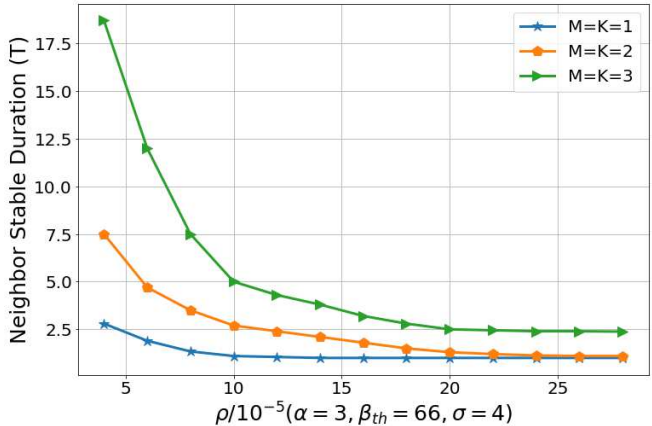}
\centering{\caption{The neighbor stable durations of node A (without prediction)}\label{}}
\end{figure}

Fig 13 and Fig 14 show the neighbor stable durations for different node densities respectively. Fig 13 shows the case of using the LQ prediction model. From the figures, it can be seen that the neighbor stable duration gradually decreases and stabilizes as the node density increases. Overall, the neighbor stable duration improves significantly with the use of the LQ prediction model compared to the case without the LQ prediction model. And as M and K increase, the neighbor stable duration also increases rapidly. Choosing the right values of M and K and using the LQ prediction model can improve the stability of the network very well.

\begin{figure}[h]
\centering
\vspace{0.2cm}
\includegraphics[width=3.4 in]{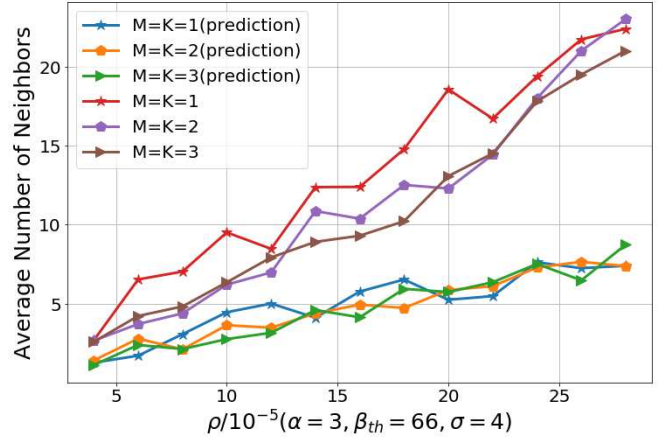}
\centering{\caption{The average number of neighbors of node A}\label{}}
\end{figure}

Although the prediction model is able to improve the neighbor stable duration well at the cost of making the effective coverage of the nodes smaller, it also excludes unstable connections at the edge of the coverage and improves the stability of the network. Fig 15 shows the change in the number of neighbors before and after the use of the LQ prediction model. As can be seen from the figure, the number of neighbors of the nodes is relatively small after the use of the LQ prediction model.

\subsection{Analysis of global topology}

In this section, we analyze the performance of the global topology in two main ways. First, the average hop count of multi-hop routes formed by any two nodes is analyzed. Then the stability of the multi-hop routes between any two nodes is analyzed. Here, we use the cumulative delivery rate of multiple wireless links in a multi-hop route, which is also called the end-to-end delivery rate, to measure the stability of a multi-hop route.

\begin{figure}[h]
\centering
\vspace{0.2cm}
\includegraphics[width=3.4 in]{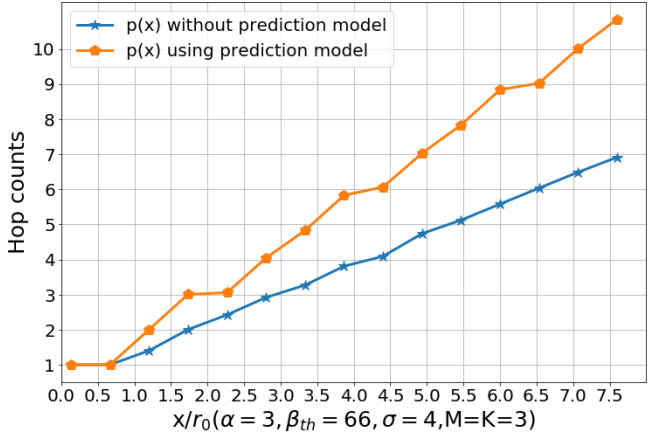}
\centering{\caption{The hop count between two given nodes}\label{}}
\end{figure}

Fig 16 shows the hop count between two nodes with different distance before and after applying the prediction model. As can be seen from the figure, the hop count increases as the distance increases and the increasing trend is faster in the network with the LQ prediction model. This is because with the use of the LQ prediction model, the unstable area of the wireless link is reduced, as shown in Fig 8. At this point, the distance of a single hop is reduced and the hop count between two nodes increases with the same distance.

\begin{figure}[h]
\centering
\vspace{0.2cm}
\includegraphics[width=3.4 in]{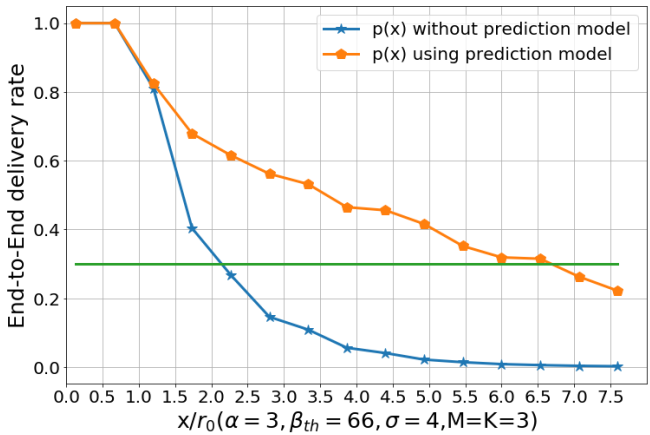}
\centering{\caption{The end-to-end delivery rate between two given nodes}\label{}}
\end{figure}

To investigate the stability of wireless multi-hop routes before and after the use of the LQ prediction model, we analyzed the end-to-end delivery rate of multi-hop routes. Fig 17 shows the end-to-end delivery rate between two nodes before and after applying the LQ prediction model. The figure shows that the end-to-end delivery rate decreases rapidly as the distance increases. In the network with the LQ prediction model, the end-to-end delivery rate for the same distance remains high despite the increase in the hop count of multi-hop routes. In the network without the LQ prediction model, the end-to-end delivery rate drops to an unacceptable level when the distance increases to a certain point, severely affecting the stability of the network. This is because, by applying the LQ prediction model, the use of unstable links is avoided. Although the hop count of multi-hop routes is increased to some extent and packets are forwarded more times, the reliability of each link segment is enhanced and the end-to-end delivery rate decreases more slowly.

If we reduce the end-to-end delivery rate to $30\%$ as the network stability threshold, we can see that in the network without the predictive model, the network radius is approximately $2.1r_0$, whereas in the network using the LQ prediction model, the network radius increases to $6.7r_0$, despite the reduced coverage of individual nodes.

\section{Conclusion}
In this paper, we delve into the role of wireless link quality prediction models in improving link stability and analyze the suppression effect of prediction models on unstable links. And we propose a software-defined wireless sensor network architecture based on the link quality prediction model, in which a link quality prediction mechanism is incorporated into the software-defined wireless sensor network architecture. We build a Python-based simulation platform for software-defined wireless sensor networks to analyze the effect of the link quality prediction model on the topology stability performance. The validation results of the simulation platform show that the usage of LQ prediction model can well avoid the application of unstable links, reduce the instability of the network and improve the overall performance of the network.

In the further research, we focus on optimization of the wireless link quality prediction model and the design of software-defined wireless sensor network protocol, and try to introduce reinforcement learning to improve the generalizability of the link prediction model.

\section*{Conflict of Interests}
The authors declare that there is no conflict of interest regarding the publication of this article.

\section*{Acknowledgment}

This work is supported by Information Engineering College Shaoguan University.



\end{document}